\documentstyle[12pt]{article}

\catcode`\@=11
\long\def\@makefntext#1{ 
\protect\noindent \hbox to 3.2pt {\hskip-.9pt
$^{{\ninerm\@thefnmark}}$\hfil}#1\hfill} 

\def\thefootnote{\fnsymbol{footnote}}
 \def\@makefnmark{\hbox to 0pt{$^{\@thefnmark}$\hss}}

\def\ps@myheadings{\let\@mkboth\@gobbletwo
\def\@oddhead{\hbox{} 
\rightmark\hfil\ninerm\thepage}
\def\@oddfoot{}\def\@evenhead{\ninerm\thepage\hfil 
\leftmark\hbox{}}\def\@evenfoot{}
\def\sectionmark##1{}\def\subsectionmark##1{}}

\begin{document}

\newcommand{\symbolfootnote}{\renewcommand{\thefootnote}
        {\fnsymbol{footnote}}}
\renewcommand{\thefootnote}{\fnsymbol{footnote}}
\newcommand{\alphfootnote}
        {\setcounter{footnote}{0}

\renewcommand{\thefootnote}{\sevenrm\alph{footnote}}}

\newcounter{sectionc}\newcounter{subsectionc}
\newcounter{subsubsectionc}
\renewcommand{\section}[1]
{\vspace{0.6cm}\addtocounter{sectionc}{1}
\setcounter{subsectionc}{0}\setcounter{subsubsectionc}{0}
\noindent
        {\bf\thesectionc. #1}\par\vspace{0.4cm}}
\renewcommand{\subsection}[1]
{\vspace{0.6cm}\addtocounter{subsectionc}{1}
        \setcounter{subsubsectionc}{0}\noindent
        {\it\thesectionc.\thesubsectionc.#1}
\par\vspace{0.4cm}}
\renewcommand{\subsubsection}[1]
{\vspace{0.6cm}\addtocounter{subsubsectionc}{1}
        \noindent
{\rm\thesectionc.\thesubsectionc.\thesubsubsectionc. #1}
\par\vspace{0.4cm}}
\newcommand{\nonumsection}[1] {\vspace{0.6cm}\noindent{\bf
#1}
        \par\vspace{0.4cm}}

\newcounter{appendixc}
\newcounter{subappendixc}[appendixc]
\newcounter{subsubappendixc}[subappendixc]
\renewcommand{\thesubappendixc}
{\Alph{appendixc}.\arabic{subappendixc}}
\renewcommand{\thesubsubappendixc}

{\Alph{appendixc}.\arabic{subappendixc}.
\arabic{subsubappendixc}}

\renewcommand{\appendix}[1] {\vspace{0.6cm}
        \refstepcounter{appendixc}
        \setcounter{figure}{0}
        \setcounter{table}{0}
        \setcounter{equation}{0}

\renewcommand{\thefigure}{\Alph{appendixc}.\arabic{figure}}

\renewcommand{\thetable}{\Alph{appendixc}.\arabic{table}}
        \renewcommand{\theappendixc}{\Alph{appendixc}}

\renewcommand{\theequation}{\Alph{appendixc}.\arabic{
equation}}
        \noindent{\bf Appendix \theappendixc.
#1}\par\vspace{0.4cm}
\par\vspace{0.4cm}}
\newcommand{\subappendix}[1] {\vspace{0.6cm}
        \refstepcounter{subappendixc}
        \noindent{\bf Appendix \thesubappendixc.#1}
\par\vspace{0.4cm}}
\newcommand{\subsubappendix}[1] {\vspace{0.6cm}
        \refstepcounter{subsubappendixc}
        \noindent{\it Appendix \thesubsubappendixc. #1}
        \par\vspace{0.4cm}}

\def\abstracts#1{{

\centering{\begin{minipage}{30pc}\tenrm\baselineskip=12pt
\noindent
        \centerline{\tenrm ABSTRACT}\vspace{0.3cm}
        \parindent=0pt #1
        \end{minipage} }\par}}

\newcommand{\bibit}{\it}
\newcommand{\bibbf}{\bf}
\renewenvironment{thebibliography}[1]
        {\begin{list}{\arabic{enumi}.}
        {\usecounter{enumi}\setlength{\parsep}{0pt}
\setlength{\leftmargin 1.25cm}{\rightmargin 0pt}
         \setlength{\itemsep}{0pt} \settowidth
        {\labelwidth}{#1.}\sloppy}}{\end{list}}

\topsep=0in\parsep=0in\itemsep=0in
\parindent=1.5pc

\newcounter{itemlistc}
\newcounter{romanlistc}
\newcounter{alphlistc}
\newcounter{arabiclistc}
\newenvironment{itemlist}
        {\setcounter{itemlistc}{0}
         \begin{list}{$\bullet$}
        {\usecounter{itemlistc}
         \setlength{\parsep}{0pt}
         \setlength{\itemsep}{0pt}}}{\end{list}}

\newenvironment{romanlist}
        {\setcounter{romanlistc}{0}
         \begin{list}{$($\roman{romanlistc}$)$}
        {\usecounter{romanlistc}
         \setlength{\parsep}{0pt}
         \setlength{\itemsep}{0pt}}}{\end{list}}

\newenvironment{alphlist}
        {\setcounter{alphlistc}{0}
         \begin{list}{$($\alph{alphlistc}$)$}
        {\usecounter{alphlistc}
         \setlength{\parsep}{0pt}
         \setlength{\itemsep}{0pt}}}{\end{list}}

\newenvironment{arabiclist}
        {\setcounter{arabiclistc}{0}
         \begin{list}{\arabic{arabiclistc}}
        {\usecounter{arabiclistc}
         \setlength{\parsep}{0pt}
         \setlength{\itemsep}{0pt}}}{\end{list}}

\newcommand{\fcaption}[1]{
        \refstepcounter{figure}
        \setbox\@tempboxa = \hbox{\tenrm Fig.~\thefigure.#1}
        \ifdim \wd\@tempboxa > 6in
           {\begin{center}
        \parbox{6in}{\tenrm\baselineskip=12pt
Fig.~\thefigure. #1 }
            \end{center}}
        \else
             {\begin{center}
             {\tenrm Fig.~\thefigure. #1}
              \end{center}}
        \fi}

\newcommand{\tcaption}[1]{
        \refstepcounter{table}
        \setbox\@tempboxa = \hbox{\tenrm Table~\thetable.#1}
        \ifdim \wd\@tempboxa > 6in
           {\begin{center}
        \parbox{6in}{\tenrm\baselineskip=12pt
Table~\thetable. #1 }
            \end{center}}
        \else
             {\begin{center}
             {\tenrm Table~\thetable. #1}
              \end{center}}
        \fi}

\def\@citex[#1]#2{\if@filesw\immediate\write\@auxout
        {\string\citation{#2}}\fi
\def\@citea{}\@cite{\@for\@citeb:=#2\do
        {\@citea\def\@citea{,}\@ifundefined
        {b@\@citeb}{{\bf ?}\@warning
        {Citation `\@citeb' on page \thepage \space
undefined}}
        {\csname b@\@citeb\endcsname}}}{#1}}

\newif\if@cghi
\def\cite{\@cghitrue\@ifnextchar [{\@tempswatrue
        \@citex}{\@tempswafalse\@citex[]}}
\def\citelow{\@cghifalse\@ifnextchar [{\@tempswatrue
        \@citex}{\@tempswafalse\@citex[]}}
\def\@cite#1#2{{$\null^{#1}$\if@tempswa\typeout
        {IJCGA warning: optional citation argument
        ignored: `#2'} \fi}}
\newcommand{\citeup}{\cite}

\def\fnm#1{$^{\mbox{\scriptsize #1}}$}
\def\fnt#1#2{\footnotetext{\kern-.3em
        {$^{\mbox{\sevenrm #1}}$}{#2}}}

\font\twelvebf=cmbx10 scaled\magstep 1
\font\twelverm=cmr10 scaled\magstep 1
\font\twelveit=cmti10 scaled\magstep 1
\font\elevenbfit=cmbxti10 scaled\magstephalf
\font\elevenbf=cmbx10 scaled\magstephalf
\font\elevenrm=cmr10 scaled\magstephalf
\font\elevenit=cmti10 scaled\magstephalf
\font\bfit=cmbxti10
\font\tenbf=cmbx10
\font\tenrm=cmr10
\font\tenit=cmti10
\font\ninebf=cmbx9
\font\ninerm=cmr9
\font\nineit=cmti9
\font\eightbf=cmbx8
\font\eightrm=cmr8
\font\eightit=cmti8

\rightline{UAHEP-9311}
\centerline{\tenbf BLACK HOLES AS P-BRANES${}^*$}
\footnotetext{${}^*$ To appear in the Proceedings of the 3rd
Workshop on Thermal Field Theories and their Applications,
Banff, Alberta, Canada, August 15-27, 1993}
\vspace{0.8cm}
\centerline{\tenrm BENJAMIN C. HARMS}
\baselineskip=13pt
\vspace{0.3cm}
\centerline{\tenrm and}
\vspace{0.3cm}
\centerline{\tenrm YVAN LEBLANC}
\baselineskip=13pt
\centerline{\tenit Department of Physics and Astronomy, The
University of Alabama}
\baselineskip=13pt
\centerline{Tuscaloosa, AL 35487-0324, USA}
\vspace{0.9cm}
\abstracts{We review briefly the thermodynamical
interpretation of black hole physics and discuss the
problems and inconsistencies in this approach.  We provide
an alternative interpretation of black holes as quantum
objects and investigate the statistical mechanics of a gas
of such objects in the microcanonical ensemble.  We argue
that the theory of black holes has the conformal properties
of duality and satisfaction of the statistical
bootstrap condition.   We show in the context of mean field
theory that the thermal vacuum is the false vacuum for a
black hole and define a microcanonical vacuum which leads to
a number density characteristic of pure states for the
Hawking radiation.  }

\vfil
\twelverm   
\baselineskip=14pt

\section{Introduction}

Nearly twenty years ago, Hawking proposed that the laws of
quantum mechanics do not hold in the creation and subsequent
evaporation of black holes\cite{hawk1} .  In this picture,
which
is based on the premise that black holes can be treated as
thermodynamical systems\cite{bek}, black holes can radiate
by the process of particles tunneling quantum mechanically
through the horizon.  The laws of quantum mechanics are
violated because the emerging radiation always has a number
density function which is characteristic of mixed states,
while the accreted radiation may have been in pure
states\cite{hawk2}.
This violates the unitarity principle.  During the period
since Hawking's original proposal,
many papers have been written on the so-called information
loss paradox which occurs in the thermodynamical
interpretation of black holes \footnote{For a review of the
subject see, for example, Ref.[4]}.

In a series of papers\cite{hl1,hl2,hl3,hl4,hl5,hl6,hl7} we
have pointed some
inconsistencies in the thermodynamical interpretation of
processes involving black holes and have offered an
alternative description of black holes.  In our description
black holes are considered to be quantum objects,
specifically, extended quantum objects or p-branes.  The
main purpose of the present work is to summarize our results
to date and to suggest possible lines of research which
follow from these results.

We begin in section {\bf 2.} with a brief summary of the
issues in
the thermodynamical interpretation of processes involving
black holes.  We discuss the inconsistencies alluded to
above in detail.  In section {\bf 3.} we discuss our
interpretation of the WKB formula as the quantum tunneling
probability and review our results for the statistical
mechanics of a gas of black holes.  In section {\bf 4.} we
discuss
the thermodynamical interpretation of black holes within the
context of mean field theory and prove that the thermal
vacuum is the false vacuum for a black hole system.  In
section {\bf 5.} we present an alternative vacuum for such a
system
and prove that the particle number density for the radiated
particles derived from this formulation represents a pure
state.  In the final section we discuss our conclusions and
future extensions of our work.

\section{Thermodynamical Interpretation of Black Holes}

Bekenstein\cite{bek} was the first to suggest that the area
of a 4-dimensional black hole can be identified with its
entropy and the
surface gravity with its temperature.  The area of a
classical black hole increases when matter falls into the
black hole.  Thus as a classical black hole accretes matter,
its entropy increases.

Bardeen, Carter and Hawking\cite{bard} derived a relation
between the mass difference of neighboring equilibrium
states of a black hole and the change in its area
\begin{eqnarray}
\Delta M = \kappa \Delta A \; ,
\end{eqnarray}
and showed that the temperature is
related to the surface gravity by
\begin{eqnarray}
T = {\kappa\over{2\pi}} \; .
\end{eqnarray}

By taking quantum effects into account, Hawking\cite{hawk3},
using an operator formalism and Gibbons and
Hawking\cite{gibb}, using the
WKB approximation, demonstrated that black holes can
radiate.  In the WKB
approximation a conical singularity develops in the
Euclidean spacetime and is removed by requiring that the
imaginary time variable be periodic with period $8\pi M$,
$M$ being the mass of the black hole.  This value of the
imaginary time is identified with the inverse temperature
$\beta_H$.  The partition function for the black hole is
given by
\begin{eqnarray}
Z &=& {\rm Tr} e^{-\beta H} \nonumber \\
&\sim & e^{-S_E} \; ,
\end{eqnarray}
where $S_E$ is the classical Euclidean action.
The Hawking entropy is given by
\begin{eqnarray}
S_H &=& \beta M - S_E  \nonumber \\
    &=& S_E \; ; \ \ (D = 4) \; .
\end{eqnarray}
A calculation of the canonical (inverse)
temperature $\beta$ shows that
\begin{eqnarray}
\beta = {\partial S_H\over{\partial E}} = \beta_H = 8\pi M
\; .
\end{eqnarray}
The Hawking entropy is determined from the
Euclidean spacetime metric.  It is related to the area of
the black hole by,
\begin{eqnarray}
S_H &=& {A\over{4}} \; .
\end{eqnarray}
For a 4-dimensional Schwarzschild black hole the radius of
the horizon is
$r_+ = 2 M$, so that
\begin{eqnarray}
S_H(M) = 4\pi M^2 = {1\over{16\pi}}\beta_H^2\; .
\end{eqnarray}

This interpretation of black holes has severe problems.  The
first problem is that the canonical specific heat, an
intrinsically positive quantity, turns out to be negative
\begin{eqnarray}
C = {\partial E\over{\partial T}} = -{\beta^2\over{8\pi}}
\end{eqnarray}
A second problem is that the partition function as
calculated the microcanonical density of states
\begin{eqnarray}
Z = \int e^{-\beta E} \Omega_H dE \; ,
\end{eqnarray}
where
\begin{eqnarray}
\Omega_H = e^{S_H} \; ,
\end{eqnarray}
blows up for all temperatures,
\begin{eqnarray}
Z = \int_0^{\infty} dE e^{-\beta E + 4\pi E^2} \; \to \infty
\; ,
\end{eqnarray}
indicating a breakdown of the WKB approximation and
therefore the inequivalence of the canonical and
microcanonical ensembles for black holes.
A third problem is of a quantum mechanical nature.  The
radiation coming out of the black hole has been shown to
have a Planckian distribution
\begin{eqnarray}
n = {1\over{e^{M\beta}-1}} \; .
\end{eqnarray}
This implies a loss of coherence, because pure states may
come into the black hole, but only mixed states come out.
This is a statement of the so-called information loss
paradox.  In the thermodynamical interpretation unitarity is
lost and with it one of the basic principles of quantum
mechanics.  This interpretation requires the abandonment of
quantum mechanics and the postulation of new physical laws.
The nature of these new laws is unknown at present.

\section{Quantum Interpretation of Black Holes}

The problems with the thermodynamical interpretation of
black holes can be avoided by adopting a different
interpretation of the WKB formula.  If the WKB formula is
interpreted to be, as is usual, the tunneling probability
per unit volume for
a particle to tunnel through the horizon of a black hole
\begin{eqnarray}
P \propto e^{-S_E(M)} \; ,
\end{eqnarray}
then the black hole can be viewed as the quantum excitation
mode of a ${D-2\over{D-4}}$-brane.  To see this , one notes
that the quantum degeneracy of states
for such an object is essentially the inverse of the
tunneling probability
\begin{eqnarray}
\sigma(M) \simeq c e^{S_E(M)} \; ,
\end{eqnarray}
where c is a constant which is determined by quantum field
theoretical corrections.

As an example we consider the Schwarzschild black hole,
which in $D$-dimensions has a Euclidean metric given by
\begin{eqnarray}
ds^2 = e^{2\Phi} d\tau^2 + e^{-2\Phi} dr^2 + r^2 d\Omega_{D-
2}^2 \; ,
\end{eqnarray}
where
\begin{eqnarray}
e^{2\Phi} = 1 - \Bigl({r_+\over{r}}\Bigr)^{D-3} \; .
\end{eqnarray}
The Euclidean action calculated from this metric is
\begin{eqnarray}
S_E = {A_{D-2}\over{16\pi}} \beta_H r_+^{D-3},
\end{eqnarray}
with
\begin{eqnarray}
\beta_H = {2\pi\over{[e^{\Phi}\partial_r e^{\Phi}]_{r=r_+}}}
= {4\pi r_+ \over{D-3}} \nonumber \\
M = {D-2\over{16\pi}} A_{D-2} r_+^{D-2} \; ,
\end{eqnarray}
where $A_{D-2}$ is the area of a unit $D-2$ sphere.
Eliminating the horizon radius $r_+$ in favor of the mass,
the Euclidean action becomes
\begin{eqnarray}
S_E = C(D) M^{D-2\over{D-3}} \; ,
\end{eqnarray}
where $C(D)$ is the dimension-dependent constant
\begin{eqnarray}
C(D) = {4^{D-1\over{D-3}} \pi^{D-2\over{D-3}} \over{(D-3)(D-
2)^{D-2\over{D-3}} A^{1\over{D-3}}_{D-2}}} \; .
\end{eqnarray}
Substituting in for $S_E$ in the quantum degeneracy of
states expression we find
\begin{eqnarray}
\sigma(M) \simeq c e^{C(D)M^{D-2/D-3}} \; .
\end{eqnarray}
Comparing this expression to those known for nonlocal field
theories\cite{fubi,deth,stru,alva}, we find that it
corresponds to the degeneracy of
states for an extended quantum object ($p$-brane) of
dimension $p = {D-2\over{D-4}}$.

Returning to the thermodynamical interpretation, we find for
the statistical mechanical density of states
\begin{eqnarray}
\Omega_H(M) = e^{S_H(M)} \;
\end{eqnarray}
where $S_H$ is the Hawking entropy
\begin{eqnarray}
S_H = \beta_H M - \beta_H F(\beta_H) = \beta_H M - S_E \; .
\end{eqnarray}
In $D$-dimensions we find from Eq.(18) and Eq.(19) that the
entropy is
\begin{eqnarray}
S_H = (D-3) S_E(M) \; ,
\end{eqnarray}
giving for the density of states
\begin{eqnarray}
\Omega_H(M) \simeq \sigma^{D-3}(M) \; .
\end{eqnarray}
The partition function obtained from this density of states
is
\begin{eqnarray}
Z(\beta) &=& \int_0^{\infty} dE e^{-\beta E}\; e^{(D-
3)C(D)E^{D-2\over{D-3}}} \\
&&\to \infty \; {\rm for}\; D \ge 4 \; , \nonumber
\end{eqnarray}
illustrating for $D$-dimensional black holes the earlier
results for 4-dimensional black holes.  The statistical
mechanics of black holes must therefore be studied in the
more fundamental microcanonical ensemble.

Within the context of our interpretation of black holes as
quantum objects we have considered a gas of such objects.
The microcanonical ensemble in 4 dimensions is determined by
the microcanonical densities
\begin{eqnarray}
\Omega(E,V) = \sum_{n=1}^{\infty}\Omega_n(E,V) \; ,
\end{eqnarray}
with
\begin{eqnarray}
\Omega_n(E,V) &=& \Bigl[{V\over{(2\pi)^3}}\Bigr]^n
{1\over{n!}} \prod_{i=1}^n\Bigl( \int_{m_0}^{\infty} dm_i
\sigma_{BH}(m_i) \int_{-\infty}^{\infty} d^3p_i\Bigr)
\nonumber \\
&&\times \delta\Bigl(E - \sum_{i=1}^n E_i\Bigr)
\delta^3\Bigl(\sum_{i=1}^n p_i\Bigr) \; ,
\end{eqnarray}
where $\sigma_{BH}(m_i)$ is given by Eq.(21) and $m_0$ is
the mass of an extreme black hole ($m_0 = 0$ for
Schwarzschild black holes).  At high
energy $E$,
\begin{eqnarray}
\Omega_n \simeq \Bigl[{c V\over{(2\pi)^3}}\Bigr]^n
{1\over{n!}} e^{4\pi[E-(n-1)m_0]^2} e^{4\pi(n-1)m_0^2} \; .
\end{eqnarray}
The most probable equilibrium configuration is determined
from the condition
\begin{eqnarray}
{d\Omega_n(E,V)\over{dn}}\Big|_{n=N(E,V)} = 0 \; .
\end{eqnarray}
The equilibrium state found from this condition is very
inhomogeneous.  For $N$ black holes there are one massive
and $N-1$
massless black holes in the gas (Fig.1).
\begin{figure}
\vspace{3.5in}
\fcaption{The equilibrium configuration of $N$ black holes
(upper right)
is one massive black hole (circled cross) and $N-
1$ massless black holes. By the bootstrap property, the gas
provides a statistical model for a single quantum black
hole.}
\end{figure}
The entropy
for such a gas can be approximated by
\begin{eqnarray}
S(E,V) &\equiv& \ln \Omega(E,V) \simeq \ln
\Omega_N(E,V)\nonumber \\
&\simeq& N\ln\Bigl[{c V\over{(2\pi)^3}}\Bigr] - \ln
\Gamma(N+1) + S_H(E) \; ,
\end{eqnarray}
where $S_H(E)$ is the Hawking entropy.  The microcanonical
temperature is
\begin{eqnarray}
\beta = {dS\over{dE}} = {dS_H\over{dE}} = 8\pi E \; .
\end{eqnarray}
{}From this expression we see that the microcanonical
temperature is the same as the Hawking temperature of the
most massive black hole in the ensemble.  The microcanonical
specific heat turns out to be negative
\begin{eqnarray}
C_V = -\beta^2{dE\over{d\beta}} = -{\beta^2\over{8\pi}} \; .
\end{eqnarray}
This does not create a problem, however, because the
microcanonical specific heat is allowed in principle to be
negative.  Our statistical analysis of a gas of
Schwarzschild black holes has revealed that it obeys the
statistical bootstrap condition
\begin{eqnarray}
{\Omega(E)\over{\sigma(E)}} \to 1, \; \ \ E \to \infty \; .
\end{eqnarray}
Such a property is also pictured in Fig. 1.
Also we have shown\cite{hl1} that black hole scattering
amplitudes
are dual in the sense that the number of open channels grows
in parallel with the degeneracy of states as the energy is
increased.  These two properties are evidence of the
conformal nature of black holes as quantum objects.

\section{Mean Field Theory}

In addition to the WKB approximation there is another
semiclassical approximation which can be used to study the
properties of black holes.  In the mean field approximation
fields are quantized on a classical black hole background.
Since black holes have a horizon which causally separates
two regions of space, we must double the number of degrees
of freedom.  Two Fock spaces are required.  This doubling of
the number of degrees of freedom bears a strong resemblance
to thermofield dynamics (TFD)\cite{umez}.

If we look at the mean field theory, we find
that the vacuum for quantum fields scattered off of
black holes can be written as
\begin{eqnarray}
\mid out;0> = {1\over{Z^{1/2}(\beta)}} \sum_{n=0}^{\infty}
e^{-\beta n\omega/2}\mid n>\otimes \mid\tilde{n}> \; ,
\end{eqnarray}
where
\begin{eqnarray}
Z = \sum_{n=0}^{\infty} e^{-\beta n\omega} \; ,
\end{eqnarray}
and the $\mid\tilde{n}>$ states provide a basis set for the
sector causally disconnected from the observer.  Physical
operators are defined
on the basis set for the states $\mid n>$ outside the
horizon.
The expectation value of a physically observable operator
$\cal O$ in the $out$ region is given as,
\begin{eqnarray}
<out; 0\mid{\cal O}\mid out; 0> = {1\over{Z(\beta)}}\sum_n
e^{-n\beta\omega} <n\mid {\cal O}\mid n> \; ,
\end{eqnarray}
which corresponds to a canonical ensemble average.
As before the temperature is determined by the surface
gravity\cite{birr}
\begin{eqnarray}
\beta = {2\pi\over{\kappa}} = \beta_H \; .
\end{eqnarray}
If the operator $\cal O$ is chosen to be the number operator
\begin{eqnarray}
{\cal O} = a_k^{\dagger}a_k \; ,
\end{eqnarray}
then the expression in Eq.(37) is the number density
\begin{eqnarray}
n_k(m;\beta_H) = {1\over{e^{\beta_H\omega_k(m)} - 1}} \; .
\end{eqnarray}
If black holes are described by a local field theory, this
expression presents a problem because it implies loss of
coherence.  The $in$ state is a pure state
\begin{eqnarray}
\mid in;0> = \mid 0>\otimes\mid \tilde{0}> \; ,
\end{eqnarray}
but the number density obtained from the outgoing states is
a thermal distribution.

If black holes are interpreted as quantum excitations of a
p-brane, non-local effects must be taken into account.  To
take account of all possible mass states, we must sum over
the mass
\begin{eqnarray}
n_k(\beta_H) = \int_0^{\infty} dm \; \sigma(m)
n_k(m;\beta_H) \; .
\end{eqnarray}
The thermal vacuum must now be modified to include
contributions from all possible mass states,
\begin{eqnarray}
\mid out;0> = {1\over{Z^{1/2}(\beta)}}\Bigl[\prod_{m,k}\;
\sum_{n_{k,m} = 0}^{\infty}\Bigr]\prod_{m,k} e^{-\beta
n_{k,m}\omega_{k,m}/2}\mid n_{k,m}>\otimes\mid
\tilde{n}_{k,m}> \; ,
\end{eqnarray}
in which the momentum states are shown explicitly.
The quantity in square brackets represents the product of
sums over the discrete values of the momentum and mass.
Changing from discrete to continuous values of the momentum
and mass, we find for the canonical partition function
\begin{eqnarray}
Z(\beta) = \exp\Bigl({-V\over{(2\pi)^{D-1}}}\int_{-
\infty}^{\infty} d^{D-1}\vec{k} \int_0^{\infty} dm\;
\sigma(m)\ln[1 - e^{-\beta_H\omega_k(m)}]\Bigr) \; .
\end{eqnarray}
This expression can be equated to the definition of the
partition function to obtain Hagedorn's self-consistency
condition\cite{hage} for a system of particles in
thermodynamical equilibrium
\begin{eqnarray}
\int_0^{\infty}\Omega(E) e^{-\beta E} dE &=& \exp\Bigl({-
V\over{(2\pi)^{D-1}}}\int_{-\infty}^{\infty}d^{D-
1}\vec{k}\nonumber \\
&&\times \int_0^{\infty}dm\; \sigma(m)\;\ln [1 - e^{-
\beta_H\omega_k(m)}]\Bigr) \; .
\end{eqnarray}
The only objects which satisfy the self-consistency
condition are strings\cite{hage,frau,carl}
\begin{eqnarray}
\sigma(m) \sim e^{bm} \; \ \ (m \to \infty) \; ,
\end{eqnarray}
for $\beta_H > b = {\rm Hagedorn's \ \ inverse \ \
temperature}$.
Black holes do not satisfy Hagedorn's condition because
\begin{eqnarray}
\sigma_{BH}(m) \sim e^{C(D)m^{D-2/D-3}} \; ,
\end{eqnarray}
so that the exponent of m is always greater than 1 (unless
$D = \infty$).  The black hole system is not in thermal
equilibrium because it is not a self-consistent solution
under the assumption of thermal equilibrium.   We conclude
therefore that the thermal vacuum is the false vacuum for
the black hole system.

\section{Microcanonical Formulation}

To determine the true vacuum for a black hole system we
begin by writing the thermal vacuum in terms of the density
matrix $\hat{\rho}$
\begin{eqnarray}
\mid 0(\beta)> = \hat{\rho}^{1/2} (\beta,{\bf H})\mid\Im >
\; ,
\end{eqnarray}
for the case of thermal equilibrium, where
\begin{eqnarray}
\hat{\rho}(\beta,{\bf H}) &=& {\rho(\beta,{\bf
H})\over{<\Im\mid \rho(\beta,{\bf H})\mid\Im >}} \;
,\nonumber \\
\rho(\beta, {\bf H}) &=& e^{-\beta {\bf H}} \; ,\nonumber \\
\mid \Im> &=& \Bigl[\prod_{k,m} \;
\sum_{n_{k,m}=0}^{\infty}\Bigr]\prod_{k,m}\mid
n_{k,m}>\otimes\mid \tilde{n}_{k,m}> \; .
\end{eqnarray}
The power of $1\over{2}$ on the density matrix $\hat{\rho}$
is somewhat arbitrary as will be discussed below.

Observable quantities are obtained by evaluating ${\rm
Tr}\cal O$ where $\cal O$ represents any observable operator
\begin{eqnarray}
{\rm Tr}{\cal O} = <\Im\mid\cal O\mid\Im> \; .
\end{eqnarray}
In particular we consider the free field propagator, which
is defined by the relation
\begin{eqnarray}
\Delta_{\beta,\alpha}^{a\; b}(x_1,x_2) = -i<\Im\mid
T\hat{\rho}^{1-\alpha} \phi^a(x_1)
\phi^b(x_2)\hat{\rho}^{\alpha}\mid\Im> \; ,
\end{eqnarray}
where the $\phi^a$'s are the so-called thermal doublets
\begin{eqnarray}
\phi^a(x) = \left(\begin{array}{c}\phi(x) \\
\tilde{\phi}^{\dagger}(x)\end{array}\right) \; ,
\end{eqnarray}
and $\alpha$ is a parameter which can be chosen between 0
and 1.  In writing Eq.(51) use has been made of the
invariance of the trace under cyclic rotation of the
operators\cite{arim}. The Fourier transform of the
propagator in Eq.(51) is
\begin{eqnarray}
\Delta_{\beta,\alpha}^{ab} = {\tau_3\over{k^2 + m^2 -
i\epsilon\tau_3}} + {2\pi i \;
\delta(k^2+m^2)\over{e^{\beta|k_0|} -
1}}\left(\begin{array}{cc}
1 & e^{(1-\alpha)\beta |k_0|} \\ e^{\alpha\beta |k_0|} & 1
\end{array}\right) \; ,
\end{eqnarray}
where $\tau_3$ is the Pauli matrix
\begin{eqnarray}
\tau_3 = \left(\begin{array}{cc} 1 & 0 \\
0 & -1 \end{array}\right) \; .
\end{eqnarray}

In order to determine the true vacuum for this system it
will be convenient to choose $\alpha = 1$.  In this case the
thermal vacuum is given by
\begin{eqnarray}
\mid \beta> = \hat{\rho}(\beta; {\bf H})\mid \Im> \; .
\end{eqnarray}
We can now formally define the microcanonical vacuum $\mid
E>$
by the expression
\begin{eqnarray}
\mid\beta> = \int_0^{\infty} e^{-\beta E} \mid E> dE \; .
\end{eqnarray}
The microcanonical analysis of the previous sections
strongly suggests these considerations.
In this basis physical correlation functions are expressed
as
\begin{eqnarray}
G_E^{a_1...a_N}(1,2,...,N) = <\Im\mid
T\phi^{a_1}(1),...,\phi^{a_N}(N)\mid E> \; ,
\end{eqnarray}

The normalization for the microcanonical vacuum can be
determined from the the fact that the thermal vacuum
normalization is given by,
\begin{eqnarray}
<\Im\mid \beta> = 1 \; .
\end{eqnarray}
Thus according to the definition of $\mid E>$ we must have
\begin{eqnarray}
<\Im \mid E> = \delta(E) \; .
\end{eqnarray}
An explicit form for the microcanonical vacuum can be
formally obtained from the expression for the thermal vacuum
with $\alpha = 1$, Eq.(55),
by an inverse Laplace transform.  Using the fact that the
partition function obeys the relation
\begin{eqnarray}
Z^{\pm 1}(\beta,V) = \int_0^{\infty} dE\; e^{-\beta E}
\Omega(E, \pm V) \: ,
\end{eqnarray}
and the density matrix obeys the relation
\begin{eqnarray}
\rho(\beta , {\bf H}) = \int_0^{\infty} dE\; e^{-\beta
E}\rho(E,{\bf H}) \; ,
\end{eqnarray}
where the microcanonical density matrix is given by
\begin{eqnarray}
\rho(E,{\bf H}) = \delta(E-{\bf H}) \; ,
\end{eqnarray}
the microcanonical vacuum $\mid E>$ is found to be
\begin{eqnarray}
\mid E > = \Omega(E-{\bf H}, -V)\mid\Im> \; ,
\end{eqnarray}
with
\begin{eqnarray}
\Omega(E,V) &=& \delta(E) +
\sum_{n=1}^{\infty}\Bigl({V\over{(2\pi)^{D-1}}}\Bigr)^n
{1\over{n!}}\Bigl[\prod_{i=1}^n \int_0^{\infty} dm_i\;
\sigma(m) \nonumber \\
&&\times \int_{-\infty}^{\infty} d^{D-1}\vec{k}_i \sum_{l_i
= 1}^{\infty} {1\over{l_1,l_2,...,l_n}} \delta(E -
\sum_{i=1}^n l_i\omega_{k_i}(m_i))\Bigr] \; .
\end{eqnarray}
In this expression $\sigma(m_i)$ is the black hole
degeneracy of states and $\omega_k(m) = \sqrt{k^2 + m^2}$.
The first term in Eq.(64) comes from the vacuum sector $(E =
0)$.  If $E>0$, then only the second term contributes.  We
can define a normalized vacuum as
\begin{eqnarray}
\mid 0(E)> = {\mid E>\over{\delta(0)}} \; ,
\end{eqnarray}
so that
\begin{eqnarray}
<\Im\mid 0(E)> = 1 \; , \ \ (E = 0) \; .
\end{eqnarray}

The set of equations (57,59,63,64,and 65) defines our
quantum theory of fields in black hole spacetimes.
Interaction effects have been negelected up to this point,
but they can be taken into account by means of the
microcanonical propagator.  Using the expression for the
2-point function
\begin{eqnarray}
G_E^{a_1 a_2}(x_1,x_2) = <\Im\mid
T\phi^{a_1}(x_1)\phi^{a_2}(x_2)\mid 0(E)> \; ,
\end{eqnarray}
and Eq.(66), we can obtain the microcanonical propagator
\begin{eqnarray}
\Delta_{E,1}^{ab}(k) &=& {1\over{\delta(0)}}\Bigl({\tau_3
\delta(E)\over{k^2 + m^2 -i\epsilon\tau_3}} + 2\pi i
\delta(k^2 + m^2) \nonumber \\
&&\times\Bigl[\sum_{l=1}^{\infty}\delta(E -
l|k_0|)\left(\begin{array}{cc} 1 & 1 \\ 1 & 1
\end{array}\right)
+ \delta(E)\left(\begin{array}{cc} 0 & 0 \\ 1 &
0\end{array}\right)\Bigr]\Bigr)
\end{eqnarray}
Of course only the $(1,1)$-component $\Delta_{E,1}^{11}$ is
physically observable, and this component is essentially
Weldon's propagator\cite{weld}.

We can now obtain the particle number density for Hawking
radiation for this theory .  From Eq(68) we see that
\begin{eqnarray}
n_{k,m} = \sum_l {\delta(E-l\omega_k(m))\over{\delta(0)}} \;
,
\end{eqnarray}
The question is: Does this describe a pure state?  To answer
this question we note that a necessary and sufficient
condition for a density matrix to describe a pure state is
the idempotency condition
\begin{eqnarray}
\int_0^{\infty} dE' \hat{\rho}_{E_1 E'} \hat{\rho}_{E'E_2} =
\hat{\rho}_{E_1 E_2} \; .
\end{eqnarray}
In this case
\begin{eqnarray}
\hat{\rho}_{E_1 E_2}(E) = {\delta(E-E_1)\delta(E_1-
E_2)\over{\delta(0)}} \; ,
\end{eqnarray}
where
\begin{eqnarray}
\hat{\rho}(E,{\bf H}) = {\delta(E-{\bf H})\over{\delta(0)}}
\; ,
\end{eqnarray}
and
\begin{eqnarray}
\int dE \hat{\rho}_{E E} = 1 \; .
\end{eqnarray}
Substitution of the expression for $\hat{\rho}_{E_1E_2}(E)$
into the idempotency condition shows that this condition is
satisfied.

\section{Conclusions}

The foregoing analysis clearly shows that for semiclassical
quantization of fields in black hole backgrounds the
microcanonical ($\mid E>$) vacuum is the proper choice, not
the thermal vacuum ($\mid \beta>$).  A fixed energy basis
for the Hilbert space of the theory should be used instead
of the usual thermal state.  Black hole states are therefore
particle states.

Our analysis also made use of the fact that quantum gravity
must be treated
as a nonlocal quantum field theory.  In our interpretation
of black holes as quantum objects the quantum degeneracy of
states points to black holes as the excitation modes of
$p$-branes, with $p = {D-2\over{D-4}}$.  The self-consistent
treatment of black holes as quantum extended objects implies
that black holes are elementary particles.

One interesting extension of the quantum theory of fields in
black hole spactimes given above would be to work out the
path integral formulation of field quantization in the fixed
energy basis.  A further extension is the application of
this formalism to string theory.  We are also interested in
extracting predictions of measurable effects using the
thermodoublet formalism of black holes.

\section{Acknowledgements}
This work was supported in part by the Department of Energy
under Grant No.  DE-FG05-84ER40141.

\end{document}